\documentclass[a4paper, 12pt]{article}
\usepackage[height=22.5cm, width=16.5cm]{geometry}
\usepackage{graphicx}
\usepackage{color}
\usepackage{amsmath, amssymb}
\usepackage{wrapfig}
\usepackage{accents}

\usepackage{amsthm}

\usepackage{latexsym}

\numberwithin{equation}{section}

\makeatletter
\def\segment#1(#2)#3({\@segment(#2)(}
\def\@segment(#1, #2)(#3, #4){%
	\@tempdima#1\p@ \advance\@tempdima#3\p@
	\divide\@tempdima\tw@
	\@tempdimb#2\p@ \advance\@tempdimb#4\p@
	\divide\@tempdimb\tw@
	\edef\@segment@temp{\noexpand\qbezier(#1, #2)%
	(\strip@pt\@tempdima, \strip@pt\@tempdimb)(#3, #4)}%
\@segment@temp}
\def\widebar{\accentset{{\cc@style\underline{\mskip10mu}}}}
\makeatother

\DeclareMathOperator{\Tr}{Tr}

\DeclareMathOperator{\Sp}{{\rm Sp}}
\DeclareMathOperator{\SL}{{\rm SL}}
\DeclareMathOperator{\GL}{{\rm GL}}
\DeclareMathOperator{\U}{{\rm U}}

\begin{document}
\begin{titlepage}

	\begin{flushright}
		UT-13-21
	\end{flushright}
	\vskip 2cm

	\begin{center}
		{\Large \bfseries
			Notes on holonomy matrices of\\[.5em]
			hyperbolic 3-manifolds with cusps
		}

		\vskip 1.2cm

		Fumitaka Fukui
		\footnote{ffukui@hep-th.phys.s.u-tokyo.ac.jp}

		\bigskip
		\bigskip

		\begin{center}
			Department of Physics, Faculty of Science, \\
			University of Tokyo,  Bunkyo-ku, Tokyo 133-0022, Japan
		\end{center}

		\vskip 1.5cm

		\textbf{Abstract}
	\end{center}
	\medskip
	\noindent

	In this paper, we give a method to construct holonomy matrices of hyperbolic 3-manifolds
	by extending the known method of hyperbolic 2-manifolds.
	It enables us to consider hyperbolic 3-manifolds with nontrivial holonomies.
	We apply our method to an ideal tetrahedron and succeed in making the holonomies nontrivial.
	We also derive the partition function of the ideal tetrahedron with nontrivial holonomies
	by using the duality proposed by Dimofte, Gaiotto and Gukov.

	\bigskip
	\vfill

\end{titlepage}

\section{Introduction}

It is well known that Einstein-Hilbert gravity in three dimensions with a negative cosmological constant 
is equivalent to Chern-Simons theory \cite{Achucarro1986a, Witten1988a}.
In this correspondence vierbein and spin connection are combined into an $\SL(2, \mathbb{C})$ connection, 
and the Einstein equation and torsionless condition restrict
the $\SL(2, \mathbb{C})$ connection to be flat.
Because the Chern-Simons action includes only one derivative, 
analyzing Chern-Simons theory is usually easier than gravity theory.

In $AdS_3$ gravity theory there is an interesting solution which is known as  BTZ black hole \cite{Banados1992}.
In the BTZ background the space is topologically equivalent to a solid torus \cite{Banados1993}.
The boundary of the solid torus corresponds to the AdS boundary and 
the horizon lies on the core of the solid torus. 
BTZ solutions are characterized by two parameters, the mass and the angular momentum of the black hole, 
and they corresponds to the complex moduli of the boundary torus.
In the Chern-Simons theory side, these characteristics are captured by $\SL(2, \mathbb{C})$ holonomy matrices.
There is a non-contractible cycle in the solid torus and
the holonomy around the cycle encodes the geometrical data.

It is interesting to generalize BTZ solutions to spaces with more complicated topology.
There are two possible ways for generalization.
First, if we regard a solid torus as a trivial knot complement of $S^3$ like as \cite{Gukov2005},
one possible way is to consider a nontrivial knot complement of $S^3$.
For the case of a hyperbolic knot $K$, 
Chern-Simons theory defined on the knot complement $S^3 \backslash K$ was well studied in
\cite{Dimofte2011, Fuji2012a, Fuji2012, Witten1989a}.
Another direction we can take is to regard the BTZ solution as a torus whose inside is filled up,
and replace the boundary to general Riemann surfaces.
We will call such 3-manifold a ``solid Riemann surface''.
A particular difficulty occurs in the second generalizations
that the space can have trivalent vertices like pants, 
and we should handle this new feature appropriately.
In this paper we propose a partition function which is dual to the ideal tetrahedron 
with nontrivial holonomies.

The way we want to proceed in this paper is the second generalization.
To consider solid Riemann surfaces, the most exciting subject in this course is
the case that the Riemann surface is a pair of pants.
It is important because when we construct any solid Riemann surfaces by pants decomposition,
there (almost) always appear pairs of ``solid pants''.
The biggest reason is following;
In the case of a BTZ background space, a black hole exists at the core of the torus.
If we assume that there is a black hole at the core of the general Riemann surface, 
a pair of solid pants represents a fusion or a fission of black holes.

Our strategy depends on the nature of the Chern-Simons theory.
Let us consider the Chern-Simons theory defined on a 3-manifold $M$.
Since the theory is topological, the vacuum solutions are specified by
the topological invariants.
Because the e.o.m. of the Chern-Simons theory implies the flat connection 
condition, 
the most conventional invariant is a Wilson loop.
Actually from a mathematical perspective, 
rigidity theorem ensures that any hyperbolic 3-manifolds with finite volume can be uniquely determined
by the holonomy data of every cycle.
In other words, we can specify the solution if we know all Wilson loops of the non-contractible cycles in $M$. 

Next we consider a geometry of the boundary surface $\partial M$.
To our purpose we can restrict $\partial M$ to be a Riemann surface.
As will be reviewed in section 2, holonomy matrices are closely related to 
the geodesic lengths of the cycles.
Therefore if we suppose $\partial M$ has a hole and it has a finite geodesic length, 
the holonomy around the hole is no longer trivial.
Then if this cycle is contractible through $M$, there appears a paradox:
because of the flat connection condition, 
a Wilson loop is invariant under deformation of the cycle
and hence it should be trivial when the cycle is contractible.
This observation implies that there lies a line-shaped defect inside $M$
which prevents the cycle from shrinking, 
and this defect is called a cusp.
This means that if we want to consider general $M$, 
we are naturally guided to treat 3-manifolds with cusps.
However, the way cusps are made is just to consider nontrivial holonomies, 
thus we are able to probe the inside of $M$ by the surface data.

Inversely speaking, if we consider the case $\partial M$ has a hole but
the holonomies of the cycle is set to be identically trivial.
In this case there is no cusp ending at the hole and
we cannot know the existence of the cycle from Chern-Simons theory.

In this paper we study the solid pants with nontrivial holonomies.
In order to get a 3-manifold with nontrivial holonomies, 
we tackle an ideal tetrahedron as the simplest case at first.
Since for an ideal tetrahedron holonomies around the four vertices are trivial, 
our first task is to relax the holonomy conditions by examining the hyperbolic structure in detail.
The solid pants is obtained from a tetrahedron with nontrivial holonomies
by trivializing only one holonomy.
The computational method of hyperbolic structures is reviewed in section 2
and we discuss its application in section 3.
In the following section we propose a wave function or
a partition function of a DGG dual theory \cite{Dimofte2011}.

Our project aims
to analyze gravity theories of general solid Riemann surfaces 
and derive the partition functions of the gravity theories.
In the past studies  \cite{Carlip1995, Carlip1997}, 
the thermodynamics of the BTZ black holes are discussed by using the partition functions.
There, the partition functions were derived by using WZW model, 
but only semi-classical behaviors are surveyed in detail and one-loop corrections seems not so obvious.
We are studying the partition functions of solid Riemann surfaces and
preparing for the paper \cite{Fukui-preparing}
and we hope that our study will shed a light on the quantum behavior of three dimensional gravity.

While we were preparing the paper, a paper \cite{Dimofte2013}
which includes study on a tetrahedron with cusp defects has appeared.
They create cusps by using truncated ideal tetrahedra and 
collecting truncated vertices around the cusp to form small tubes.
Our approach is different from theirs in the point that we need no truncated tetrahedra, and
our method seems simpler when creating 3-manifolds with cusps.
Since our result has been obtained before  \cite{Dimofte2013} came up, 
we decided to separate our ongoing project and
concentrate to describe our method in this paper.

\section{Holonomy Matrices}

Let us review the relation between the hyperbolic structure and 
the holonomies of the manifold.
First we will start from two dimensions and
review the character of hyperbolic manifolds.
After that, we will extend the result of the two dimensional case
to hyperbolic 3-manifolds with boundaries.
Hyperbolic 2-manifolds are well studied and
our review of two dimensional case is based on  \cite{Chekhov2000, Chekhov2007}.

The result of this section will be used in the next section for our case, a tetrahedron with nontrivial holonomies.

\subsection{2d case}
Let us consider a Riemann surface $\Sigma _{g, h}$ of genus $g$ and $h$ holes.
Generally speaking, if $\Sigma _{g, h}$ has negative Euler number, i.e. $2-2g-h<0$, 
$\Sigma _{g, h}$ admits a negative constant curvature and
it can be embedded in the hyperbolic plane.
By the rigidity theorem, 
Riemann surface with hyperbolic metric is uniquely obtained as $\mathbb{H} / \Delta _{g, h}$, 
where $\mathbb{H}$ is a hyperbolic plane
\footnote{Here we assume that we use the Poincar\'e's upper plane model to realize $\mathbb{H}$.}
and
$\Delta _{g, h}$ is finitely generated subgroup of $\SL (2, \mathbb{R})$ known as Fuchsian group.
If we assume that we take an element $\gamma$ from the Fuchsian group $\Delta _{g, h}$ and
diagonalize it to $\gamma = \begin{pmatrix}
	e^l & 0\\
	0 & e^{-l}
\end{pmatrix}
$ with $l \in \mathbb{R}$.
When $\gamma$ acts on a point $(0, y_0)$ in the y axis, it is transferred to 
$(0, e^{2l}y_0)$ by $\gamma$, 
and taking a quotient $\mathbb{H} / \gamma$ means
that the geodesic passing $(0, y_0)$ and $(0, e^{2l}y_0)$ are compactified to make a cycle.
The geodesic length of the cycle is calculated as
\begin{equation}
	\begin{split}
		\int _{y_0} ^{e^{2l}y_0} \frac{dy}{y} = 2l =
		2\cosh ^{-1}\left(\frac{\Tr \gamma}{2}\right).
	\end{split}
\end{equation}
Because $\Tr$ operation is invariant under conjugate, 
geodesic length of the cycle is invariant under any choice of a representative of the conjugacy class.
In this way, if you choose a conjugacy class $[\gamma]$ from $\Delta _{g, h}$, 
there is a corresponding cycle which is created when $\mathbb{H}$ is divided by $\gamma$, 
and the geodesic length of the cycle is also determined by $[\gamma]$.

Construction of the Fuchsian group is very simple.
First, given a Riemann surface $\Sigma _{ g, h }$, 
we take an ideal triangulation over fundamental domain of $\Sigma _{g, h}$.
Next we give a parameter variable to each edge of triangulation.
Actually these variables span the coordinates of Teichm\"uller space known as shear coordinate \cite{Fock1997}.
Next we consider the cycle in $\Sigma _{g, h}$.
When we draw a cycle on the triangulated fundamental domain and we walk along the cycle, 
we will pass some edges and turn left or right in some triangles, 
and we can reconstruct the cycle from the ordered data of crossing which edges and 
turning whether left or right in which triangles.
One may find that the ordered data are enough to reconstruct the path.

Now we are ready to get the holonomy matrix of the cycle.
What we must do is encode the ordered data into $\SL (2, \mathbb{R})$ matrices as following way;
if we turned left ( or right) then we multiply the left-turning matrix $L$ (or right-turning matrix $R$) from left, 
or if we crossed the edge with parameter z then we multiply the edge-crossing matrix $X_z$ from left.
The matrices $L, R, X_z$ are defined as
\begin{equation}
	L=\begin{pmatrix} 0 & -1 \\ 1 & -1 
	\end{pmatrix}, 
	R=\begin{pmatrix} 1 & -1 \\ 1 & 0 
	\end{pmatrix}, 
	X_z=\begin{pmatrix} 0 & - e^{z/2} \\  e^{-z/2} & 0 
	\end{pmatrix}.
\end{equation}
The matrix L (or R) makes a left (or right) turn on the triangle 
whose vertices are $0, 1$ and $\infty$.
The matrix $X_z$ transports the points $0, 1, \infty$ to $\infty, -e^z, 0$ and vice versa.
\footnote{The meaning of these matrices are described better in  \cite{Chekhov2007}, 
but we just quote there result here.}

\begin{figure}[hbt]
	\begin{minipage}{0.49\textwidth}
		\begin{center}
			\includegraphics[trim=0 100 120 0, clip=true, width=0.95\textwidth]{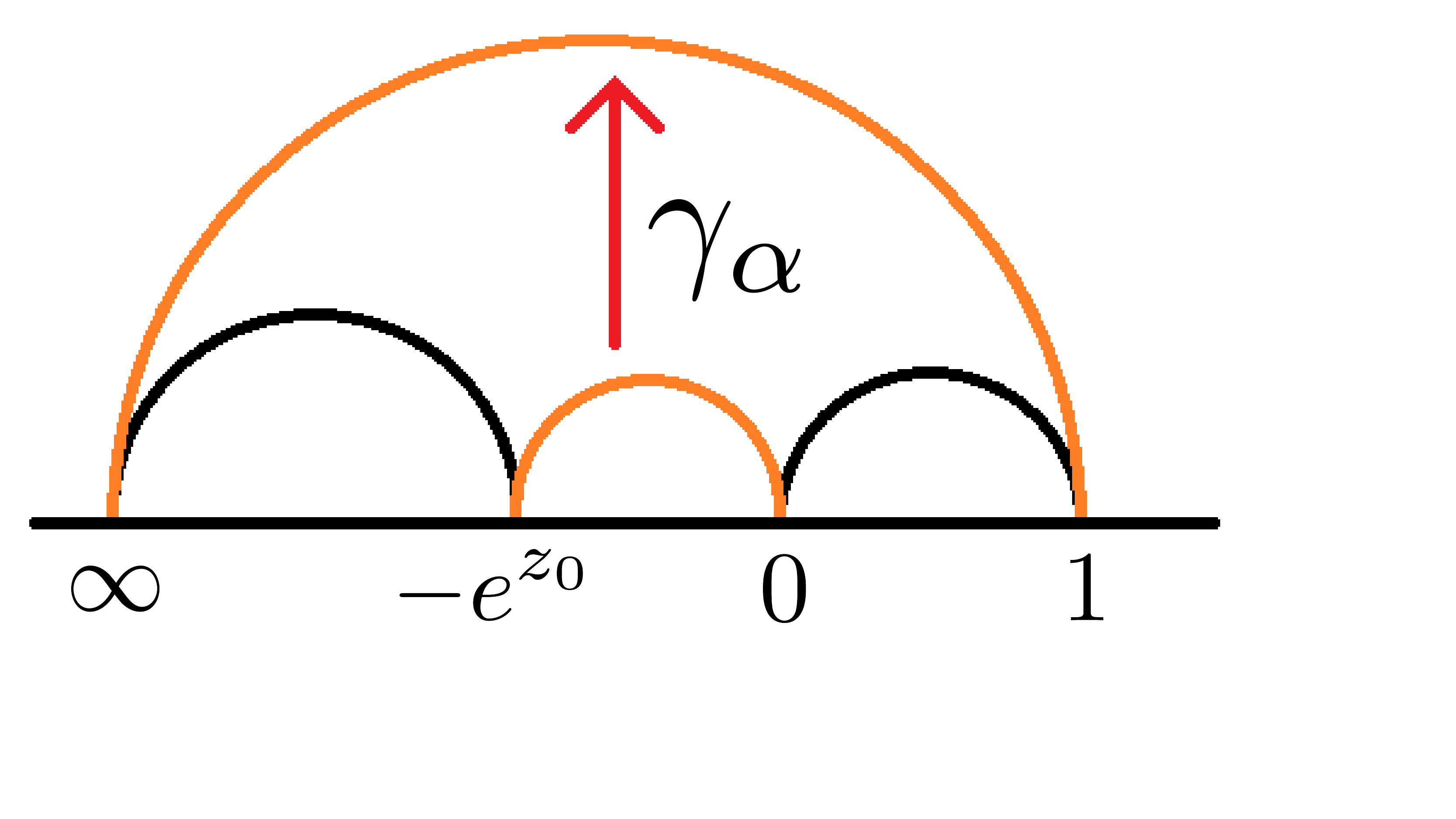}
			{(a)}
		\end{center}
	\end{minipage}
	\begin{minipage}{0.49\textwidth}
		\begin{center}
			\vspace{0.1cm}
			\includegraphics[trim=0 100 120 0, clip=true, width=0.95\textwidth]{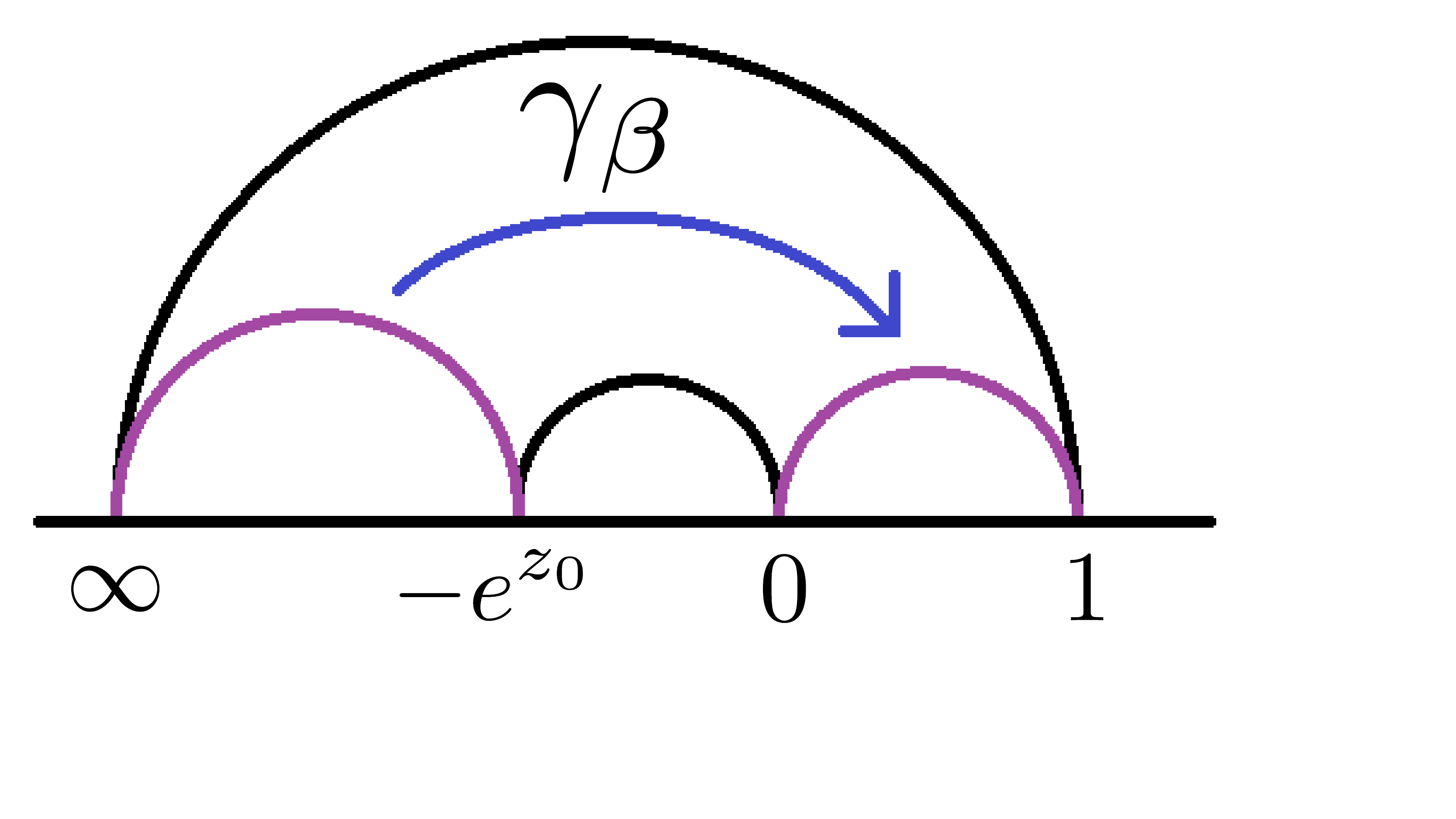}
			{(b)}
		\end{center}
	\end{minipage}
	\begin{center}
		\includegraphics[width=0.5\textwidth]{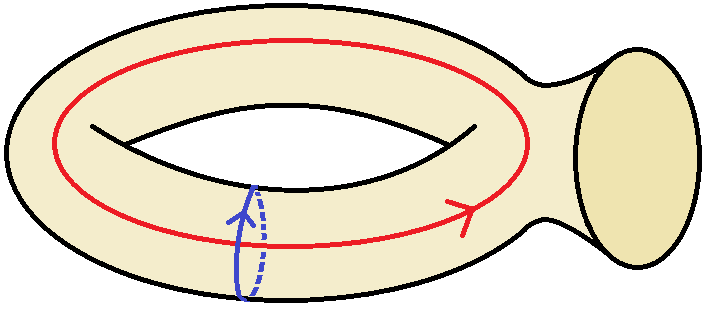} \\
	\end{center}
	\caption{We illustrate the fundamental region of $\Sigma _{1, 1}$ as a quotient
	$\mathbb{H} / \langle \gamma _\alpha , \gamma _\beta \rangle$.
	(a)The action of $\gamma _\alpha$ is depicted as a red arrow line	and the orange edges are identified.
	(b)The action of $\gamma _\beta$ is depicted as a blue arrow line and the purple edges are identified.
	(c)Two independent cycle of $\Sigma _{1, 1}$ are depicted.
	The holonomy of the red cycle is $\gamma _\alpha$ and that of the blue cycle is $\gamma _\beta$.
}
\label{fig1}
\end{figure}

We show an example how the above prescription works for a torus with one hole $\Sigma _{1, 1}$.
As in Fig.\ref{fig1}, $\Sigma _{1, 1}$ can be obtained 
as a quotient $\mathbb{H} / \langle \gamma _\alpha , \gamma _\beta \rangle$, 
where $\gamma _\alpha$ and $\gamma _\beta$ are some elements of $\SL(2, \mathbb{R})$.

By using isometry group action, we can fix three vertices of the rectangle to $0, 1, \infty$, 
and we parameterize the rest vertex to $-e^{z_0}$.
$\gamma _\alpha$ transports $0$ to $1$, $-e^{z_0}$ to $\infty$, 
and $\infty$ to somewhere in $(1, \infty)$, for instance $1+e^{-z_1}$.
Similarly $\gamma _\beta$ transports $\infty$ to $1$, $-e^{z_0}$ to $0$, 
and $0$ to somewhere in $(0, 1)$, for instance $\frac{1}{1+e^{z_2}}$.

\begin{figure}[hbt]
	\centering
	\includegraphics[trim=0 180 300 50, clip=true, width=0.7\textwidth]{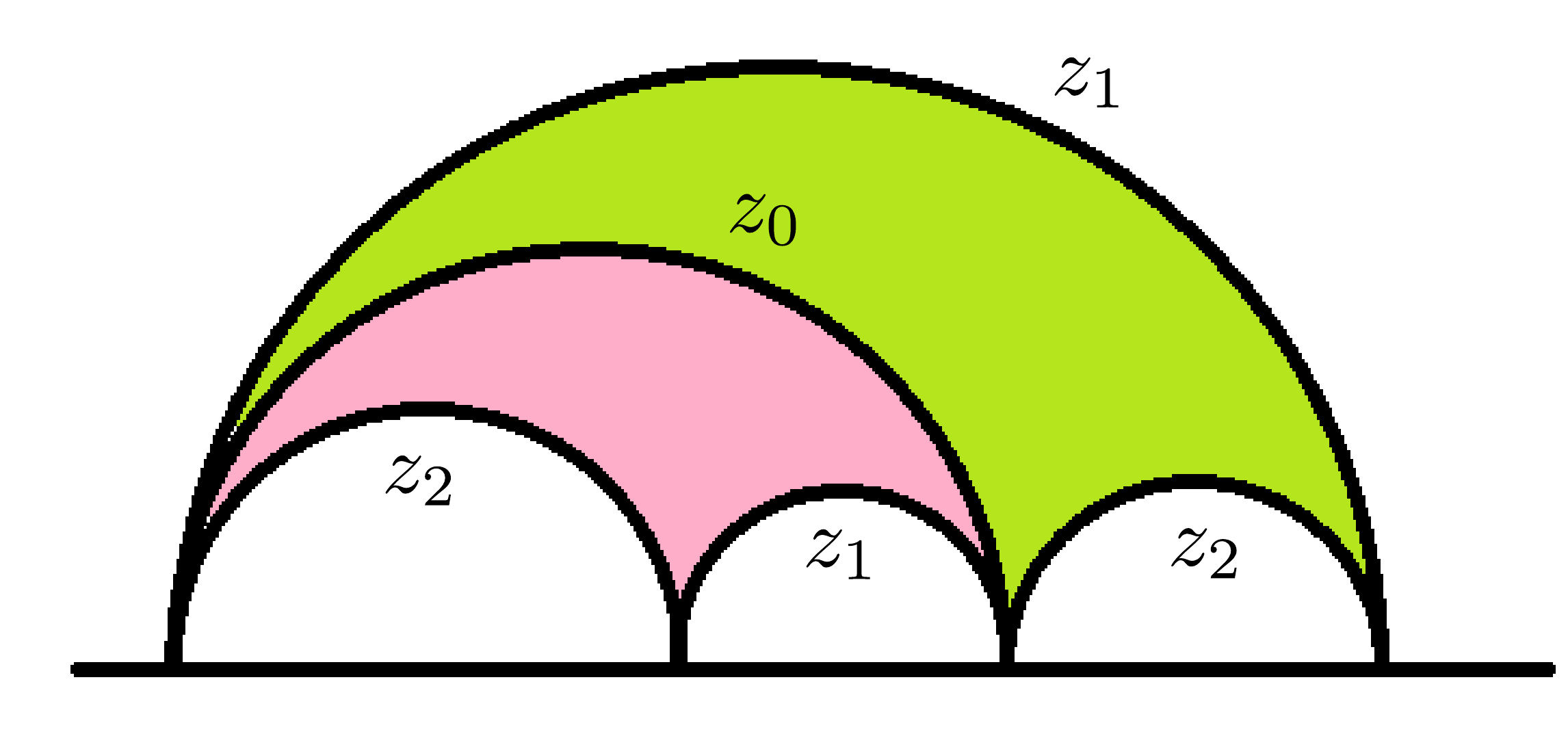}
	\caption{We make an ideal triangulation and give a parameter to each vertex.
	The identified edges have the same parameters.}
	\label{fig2}
\end{figure}

Next we make a triangulation and give parameters as \ref{fig2}.
Then $\gamma _\alpha$ can be described as starting from the pink triangle and going up, 
passing the edge with a parameter $z_0$, turning left in the green triangle, 
passing the edge with a parameter $z_1$, and turning right in the pink triangle to enclosing the path.
At last we can encode this travel into the holonomy matrix $\gamma _\alpha$ as
\begin{equation}
	\gamma _\alpha = RX_{z1}LX_{z0}.
\end{equation}
Similarly we can represent $\gamma _\beta$ as
\begin{equation}
	\gamma _\beta = LX_{z2}RX_{z0}.
\end{equation}
We can easily check that these realizations of $\gamma _\alpha$ and $\gamma _\beta$ satisfy
the expected transports.

One may think it strange that there are three parameters for moduli of a torus.
In fact these parameters include a data of a hole.
The holonomy matrices around the hole is 
$\gamma _c = \gamma _\alpha \gamma _\beta \gamma _\alpha ^{-1} \gamma _\beta ^{-1}$, 
$2\cosh ^{-1} \left( \frac{\Tr \gamma _c}{2} \right)$ is the geodesic length of the hole.
Therefore the parameters $ z_0, z_1, z_2$ correspond to the complex moduli and the moduli of the hole of 
$\Sigma _{1, 1}$.

This demonstration tells us that the essence of constructing a holonomy matrix $\gamma$ is 
to observe where $\gamma$ transports triangles.
In this example we arranged the coordinates of the vertices by hand in order that
$\gamma _\alpha$ and $\gamma _\beta$ correctly represent the transportation.
But in fact, if we define the parameters of the edges before setting the positions of the vertices, 
we can still set the positions consistent with $\gamma _\alpha$ and $\gamma _\beta$.
Thus when we compute holonomies, we only use of the parameters of edges and
the positions of vertices does not concern.

\subsection{3d case}

Just as two dimensional case, 
Rigidity theorem states that
a hyperbolic 3-manifold $M$ can be also created as a quotient space $\mathbb{H} / \Delta$
of the hyperbolic space, where $\mathbb{H}$ is a hyperbolic space here and
$\Delta$ is a subgroup of $\SL(2, \mathbb{C})$ which is an isometry group.

Particularly when $M$ has a boundary, we can apply the prescription in the previous subsection
to compute the holonomies of cycles in $\partial M$.
Because the space of the holonomies of $M$ can be derived from that of $\partial M$
by setting holonomies of contractible cycles to identity, 
holonomies of $M$ are still computable in this case.

Like as two dimensions, if $M$ can be obtained as $\mathbb{H}/ \Delta$ and
we take $\gamma \in \Delta$, 
then a cycle which corresponds to $\gamma$ does exist in $M$ and its geodesic data
are uniquely determined by $\gamma$.
So if the holonomy of some cycle of $M$ becomes nontrivial, 
the geodesic data of the cycle such as geodesic length or cusp angle also become nontrivial.
Besides, nontriviality of holonomy tells that the cycle is non-contractible
because of a cusp singularity.
Therefore, even in the case $M$ has cusp boundary inside, 
we have no need of the special treatment about the cusp, but
we only need to nontrivialize the derived holonomy matrix.

The process is almost same as in two dimensional case.
Making a triangulation of $\partial M$, 
giving parameters to vertices in the triangulation, 
decomposing cycles to the data of passing edges and turning left or right, 
and multiplying the matrix $L, R$ or $X_z$.
Here we need a slight modification as 
\begin{equation}
	R=\begin{pmatrix} 1 & -1 \\ 1 & 0 
	\end{pmatrix}, 
	L=\begin{pmatrix} 0 & -1 \\ 1 & -1 
	\end{pmatrix}, 
	X_z=\begin{pmatrix} 0 & i e^{z/2} \\ i e^{-z/2} & 0 
	\end{pmatrix}, 
\end{equation}
where $z \in \mathbb{C}$.
The change in $X_z$ is done in order to treat the orientation of 3-manifold correctly.
We will apply this construction to a tetrahedron in the next section.

\section{Holonomies of a Tetrahedron}

In this section we show how holonomy calculation works in three dimensions and
how to realize 3-manifolds with nontrivial holonomies
in a case of a tetrahedron.

\subsection{Tetrahedron with trivial holonomies}
\begin{figure}[htbp]
	\begin{tabular}[h]{cc}
		\centering
		\begin{minipage}[t]{0.45\hsize}
			\centering
			\includegraphics[width=\hsize]{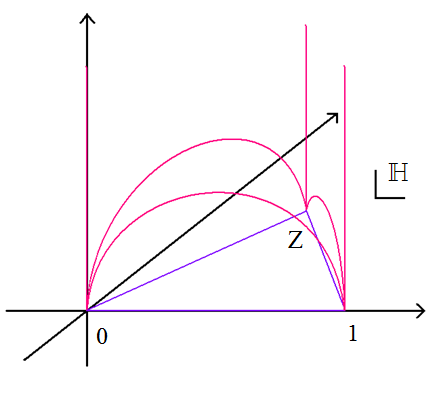}
			{(a)}
		\end{minipage}
		\begin{minipage}[t]{0.45\hsize}
			\centering
			\includegraphics[width=\hsize]{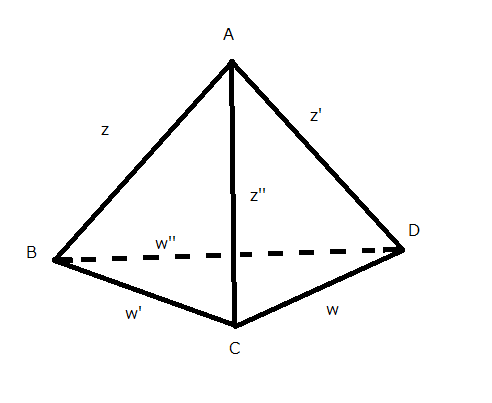}
			{(b)}
		\end{minipage}
	\end{tabular}
	\caption{(a)An illustration of an ideal tetrahedron realized in the Poincar\'e half upper space.
		Three versatile lines continue to infinity.
	(b)Making a triangulation and giving a parameter to each vertex.}
	\label{fig3}
\end{figure}
We will compute the holonomy matrices of an ideal tetrahedron.
According to the previous section, 
we should make an ideal triangulation at first.
In Fig.\ref{fig3}(b) we show an example of a triangulation 
of the surface of the ideal tetrahedron.
Next we attach parameters $z, z', z'', w, w', w''$ to the edges.

Let us see the calculation of the holonomy around the vertex A of Fig\ref{fig3}(b).
When we encircle the vertex in a counterclockwise way, 
our travel goes as
starting from the face ABD, 
passing the edge AB, turning left in the face ABC, 
passing the edge AC, turning left in the face ACD, 
passing the edge AD, and turning left in the face ABD to enclose the path.
Then we get the holonomy matrix $H_A$ around the vertex A becomes
\begin{align}
	H_A = L X_{z'} L X_{z''} L X_{z} 
	= \begin{pmatrix}
		i e ^{-\frac{1}{2}\left( z+z'+z''\right)} & 0 \\
		i e^{-\frac{1}{2}\left( z+z'+z''\right)}\left( 1-e^{z'}+e^{z'+z''}\right) 
		& - i e ^{\frac{1}{2}\left( z+z'+z''\right)} 
	\end{pmatrix}.
\end{align}
Parallelly we can obtain the holonomy matrices around  the vertices B, C, D as
\begin{eqnarray}
	H_B &= L X_{w'} L X_{w''} L X_{z} 
	&= \begin{pmatrix}
		i e ^{-\frac{1}{2}\left( z+w'+w''\right)} & 0 \\
		i e^{-\frac{1}{2}\left( z+w'+w''\right)}\left( 1-e^{w'}+e^{w'+w''}\right) 
		& - i e ^{\frac{1}{2}\left( z+w'+w''\right)} 
	\end{pmatrix}, \\
	H_C &= L X_{w'} L X_{z''} L X_{w} 
	&= \begin{pmatrix}
		i e ^{-\frac{1}{2}\left( w+w'+z''\right)} & 0 \\
		i e^{-\frac{1}{2}\left( w+w'+z''\right)}\left( 1-e^{w'}+e^{w'+z''}\right) 
		& - i e ^{\frac{1}{2}\left( w+w'+z''\right)} 
	\end{pmatrix}, \\
	H_D &= L X_{z'} L X_{z''} L X_{w} 
	&= \begin{pmatrix}
		i e ^{-\frac{1}{2}\left( w+z'+z''\right)} & 0 \\
		i e^{-\frac{1}{2}\left( w+z'+z''\right)}\left( 1-e^{z'}+e^{z'+z''}\right) 
		& - i e ^{\frac{1}{2}\left( w+z'+z''\right)} 
	\end{pmatrix}.
\end{eqnarray}
If we impose trivial holonomy conditions, the parameters of the opposite side are caused to be equal, e.g.
\begin{equation}
	\begin{split}
		z&=w \\
		z'&=w' \\
		z''&=w'' , 
	\end{split}
\end{equation}
and the rest constraints are
\begin{equation}
	\begin{split}
		z+z'+z'' &= i\pi \\
		e^{-z}+e^{z'}-1 &= 0, 
	\end{split}
\end{equation}
and permutated ones.
Therefore, under the identification of $Z=e^z, Z'=e^{z'}, Z''=e^{z''}$, 
holonomy computation goes along with a hyperbolic space.

\subsection{Tetrahedron with nontrivial holonomies}

We have seen that the holonomy calculations works well in the case of trivial holonomies, 
then how about a tetrahedron with nontrivial holonomies?

Our method needs no special treatment to non-trivializing holonomies:
just replacing the trivial holonomy conditions.
The 3-manifold we want to make is the solid pants, so
we set one of the four holonomies, $H_A$ for example, to be trivial and
the others to be nontrivial.
If we suppose $H_A = id$ and 
$\Tr H_B = 2\cosh \left( \frac{l_B}{2} \right)$ (and same for C, D), 
the relations between parameters are
\begin{equation}
	\begin{split}
		&	z + z' +z'' = i\pi \\
		&	w=z+\frac{1}{2}\left(l_B - l_C + l_D \right) \\
		&	w'=z'+\frac{1}{2}\left(-l_B + l_C + l_D \right) \\
		&	w''=z''+\frac{1}{2}\left(l_B + l_C - l_D \right) , 
	\end{split}
\end{equation}
and
\begin{equation}
	e^{-z}+e^{z'}-1=0.
\end{equation}
The parameters of the opposite edges are no longer same because of the nontriviality of holonomies.

It seems nice to represent nontrivial holonomies, but we cannot get no more conditions
over $l_B, l_C$ and $l_D.$
Actually, the holonomy matrices $H_B, H_C, H_D$ should be related to each other
because the composition of the cycles around these vertices can shrink.
To overcome this point, we should change the triangulation.

If we stick to one tetrahedron, we cannot seize holonomies at hand.
So we should divide the  tetrahedron with nontrivial holonomy into four trivial tetrahedra, 
and try to evaluate the holonomy concretely.

Our division is done like below;
\begin{figure}[htbp]
	\begin{tabular}[h]{cc}
		\centering
		\begin{minipage}[t]{0.52\hsize}
			\centering
			\includegraphics[trim=0 0 300 0, width=\hsize, clip=true]{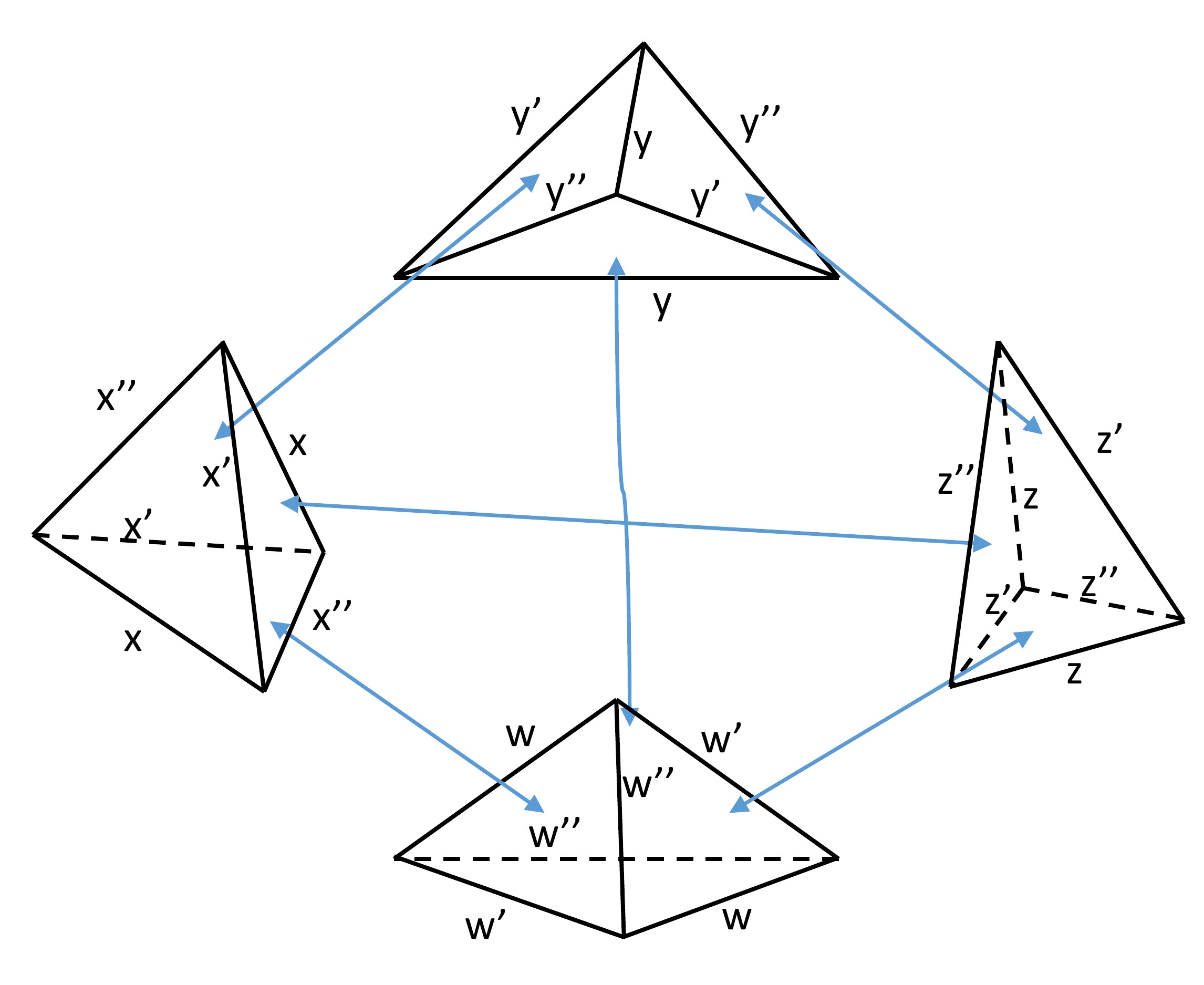}
			{(a)}
		\end{minipage}
		\begin{minipage}[t]{0.47\hsize}
			\centering
			\includegraphics[trim=0 80 380 0, width=\hsize, clip=true]{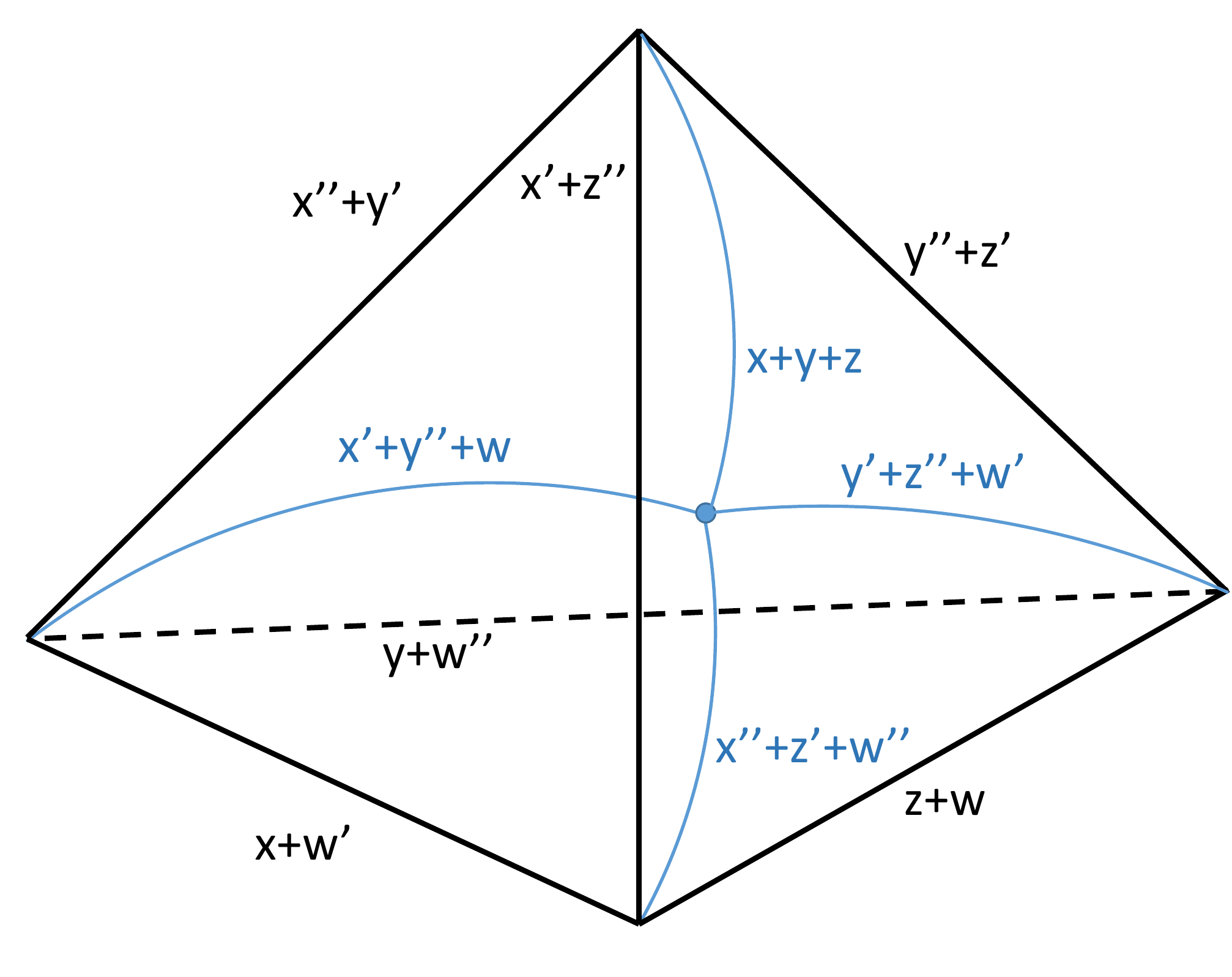}
			{(b)}
		\end{minipage}
	\end{tabular}
	\label{fig4}
	\caption{(a)The way of gluing four tetrahedra to get one tetrahedron.
		By the blue arrows the glued surfaces are indicated.
		(b)The glued tetrahedron. The blue curved lines are the internal edges and giving cusps.
	The new parameters of the new edges are written.}
\end{figure}

The parameters we have are $x, x', x'', y, y', y'', z, z', z'', w, w', w''$.
In each tetrahedron we impose trivial holonomy conditions, so we have
\begin{equation}
	\begin{split}
		x+x'+x''=i\pi \\
		e^{-x}+e^{x'}-1=0
	\end{split}
\end{equation}
and same for $y, z, w$.

Holonomy calculations go well with parameters changed.
The holonomy matrices now become
\begin{align}
	H_A &= LX_{x''+y'}LX_{y''+z'}LX_{z''+x'} \nonumber \\ 
	&=
	\begin{pmatrix}
		ie^{ -\frac{1}{2}\left(x'+x''+y'+y''+z'+z''\right)} & 0\\
		ie^{-\frac{1}{2}\left(x'+x''+y'+y''+z'+z''\right)}
		\left( 1-e^{x''+y'}+e^{x''+y'+y''+z'}\right) &
		-ie^{ \frac{1}{2}\left(x'+x''+y'+y''+z'+z''\right)}
	\end{pmatrix} \\
	H_B &= LX_{x+w'}LX_{y+w'}LX_{x''+y'} \nonumber \\
	&=
	\begin{pmatrix}
		ie^{ -\frac{1}{2}\left(w'+w''+x+x''+y+y'\right)} & 0\\
		ie^{ -\frac{1}{2}\left(w'+w''+x+x''+y+y'\right)} 
		\left( 1-e^{x+w'}+e^{x+y+w'+w''}\right) &
		-ie^{ \frac{1}{2}\left(w'+w''+x+x''+y+y'\right)}
	\end{pmatrix} \\
	H_C &= LX_{x'+z''}LX_{z+w}LX_{x+w'} \nonumber \\
	&=
	\begin{pmatrix}
		ie^{ -\frac{1}{2}\left(w+w'+x+x'+z+z''\right)} & 0\\
		ie^{ -\frac{1}{2}\left(w+w'+x+x'+z+z''\right)} 
		\left( 1-e^{x'+z''}+e^{x'+z+z''+w}\right) &
		-ie^{ \frac{1}{2}\left(w+w'+x+x'+z+z''\right)}
	\end{pmatrix} \\
	H_D &= LX_{y''+z'}LX_{y+w''}LX_{z+w} \nonumber \\
	&=
	\begin{pmatrix}
		ie^{ -\frac{1}{2}\left(w+w''+y+y''+z+z'\right)} & 0\\
		ie^{ -\frac{1}{2}\left(w+w''+y+y''+z+z'\right)} 
		\left( 1-e^{y''+z'}+e^{w'+y+y''+z'}\right) &
		-ie^{ \frac{1}{2}\left(w+w''+y+y''+z+z'\right)}
	\end{pmatrix}.
\end{align}
If we constrain these holonomy matrices to be $H_A=id, H_B=2\cosh(\frac{l_B}{2})$
and same for C, D, 
then we get the following relations between parameters;
\begin{eqnarray}
	&x+y+z &= 2\pi i \\
	&w+x'+y'' &= l_B + 2\pi i \\
	&w''+x''+z' &= l_C + 2\pi i \\
	&w'+y'+z' &= l_D + 2\pi i.
\end{eqnarray}
In this case if we sum up the both sides of these equation, 
a relation
\begin{equation}
	l_B+l_C+l_D=0
\end{equation}
can be obtained, and we can relate the holonomies $H_B, H_C, H_D$ successfully.
These relations are needed to compute the partition function in the next section.

\section{Partition functions}

In this section we will discuss the partition function of the hyperbolic 3-function $M$, 
particularly focusing on the ideal tetrahedron.
The calculations of the partition functions are well described in \cite{Dimofte2011}, 
we will shortly review this in order to make this paper self-contained.

In \cite{Terashima2011, Dimofte2011} it was pointed out that
there was a correspondence between the partition function of gravity theory
and that of $\mathcal{N}=2$ supersymmetric Chern-Simons gauge theory 
by suitably identifying parameters.
Let us come to the case of the ideal tetrahedron.
On the side of the gravity theory, we can consider the partition function of the gravity theory 
defined on the ideal tetrahedron  \cite{Chekhov1999, Chekhov2000}.
If we set the vertices at $\{0, 1, \infty, Z\}$, $Z$ becomes a parameter of the partition function.
We will denote the partition function as $Z^{\hbar}(Z)$.
On the side of dual gauge theory, 
we can consider the $\mathcal{N}=2$ gauge theory defined on $S^3_b$ \cite{Hama2011a}.
The matter content of the dual theory is one chiral matter.
The $\U(1)$ flavor symmetry is gauged and
a Chern-Simons coupling of level $-frac{1}{2}$ is added to the Lagrangian.
For later use we will call this theory $\mathcal{T}_{\Delta}$.
The parameters of the theory are the mass $m$ and the R-charge $R$ of the chiral matter, 
thus we will denote the partition function as $Z_{\Delta}(\tilde{m})$
where $ \tilde{m}=m+\frac{iQ}{2}R$.
The identification of the parameters are as below;
\begin{eqnarray}
	\tilde{m} & \leftrightarrow & \frac{Z}{2\pi b} \\
	b^2 & \leftrightarrow & 8G, 
\end{eqnarray}
where the left hand side is of the supersymmetric gauge theory and the right hand side is of the gravity theory.
$b$ is the squashing parameter of $S^3_b$, 
and $G$ is the gravitational constant.

We can exactly determine the partition function by using the localization technique \cite{Hama2011a}, 
and the partition function of $\mathcal{T}_{\Delta}$ was
\begin{equation}
	Z_{\Delta}(\tilde {m})=e^{\frac{i\pi}{2}\left(\frac{iQ}{2}-\tilde{m}\right)^2}
	s_b\left(\frac{iQ}{2}-\tilde{m}\right), 
\end{equation}
where $Q=b+b^{-1}$.
If there are multiple tetrahedra, 
the corresponding partition function is obtained as a multiples of $Z_{\Delta}$'s.

We can define the symplectic transformation over the partition functions \cite{Witten2003a, Aharony1997}.
In the case of one parameter, the generating matrices $T$ and $S$ of $\SL(2, \mathbb{Z})$
act on the partition functions as 
\begin{eqnarray}
	T: Z(\tilde{m})& \mapsto& Z'(\tilde{m})=e^{i\pi\tilde{m}^2} Z(\tilde{m}) \\
	S: Z(\tilde{m})& \mapsto& Z'(\tilde{m}')=\int d\tilde{m} e^{-2i\pi \tilde{m}\tilde{m}'} Z(\tilde{m}).
\end{eqnarray}
The integral is over $\tilde{m} \in \mathbb{R}$.
We can check that they satisfy $S^2 = -id$ and $(ST)^3=id$, as expected.
Particularly for an ideal tetrahedron, when the transformation $ST$ acts on the $Z_{\Delta}$, 
we can get 
\begin{equation}
	\begin{split}
		ST: Z_{\Delta}(\tilde{m}) \mapsto Z^{ST}_{\Delta}(\tilde{m}') 
		&=	\int d\tilde{m} e^{-2i\pi \tilde{m}\tilde{m}' -i\pi\tilde{m}^2}Z_{\Delta}(\tilde{m})  \\
		&= -e^{-\frac{i\pi}{12}\left(1+Q^2\right)}Z_{\Delta}\left(\tilde{m}'+\frac{iQ}{2}\right).
	\end{split}
\end{equation}
Here and after we use formulae about double sine functions from  \cite{Faddeev2000}, 
we arranged some formulae and collected into Appendix.
We can find that $ST$ transformation does not change the form of $Z_{\Delta}$.

Now we should concern the variable changes of moduli parameters $z \sim w''$.
In fact, Chern-Simons action induces symplectic structures to the moduli parameters
\cite{Verlinde:1989hv, Fock1998}.
For an ideal tetrahedron the dimension of the phase space is two, 
and Poisson brackets are $\{z, z''\}=\{z'', z\}=\{z'', z'\}=1$.
\footnote{When we choose the  coordinate and the momentum as $z$ and $z''$, 
the remaining parameter $z'=i\pi-z-z''$ is not an independent variable.}
The choice of coordinates and momenta is called the polarization.
The coordinate transformations which does not change the commutation relation
are allowed, and it is symplectic group symmetry.

Let's come back to our case of the tetrahedron with nontrivial holonomies.
The phase space of nontrivial tetrahedron has dimension 8 because we started from 4 tetrahedra.
When the four tetrahedra are taken apart, 
the conventional set of independent parameters is $\{ x, y, z, w\}$ 
and the polarization is
$	\left(	z'' , w , x'' , y' , z' , w'' , x' , y\right)^{\bf T}$, 
where the first four variables are the coordinate and the last four represent the momenta.
When the tetrahedra are glued, however, these parameters can be no longer independent.
A more important thing is that these parameters does not appear solely in Fig.\ref{fig3}.
Our choice of polarization is
\begin{equation}
	\begin{pmatrix}
		X \\ C_B \\ C_C \\ C_D \\ P \\ \Theta _B \\ \Theta _C \\ \Theta _D
	\end{pmatrix}
	=\begin{pmatrix}
		x'+z'' \\ w+x'+y'' \\ w''+x''+z' \\ w'+y'+z'' \\ z'+y'' \\ \Theta _B \\ \Theta _C \\\Theta _D
	\end{pmatrix}, 
\end{equation}
where the top four are coordinates and the bottom four are momenta of the phase space.
The coordinates $X, C_B, C_C, C_D$ correspond to edge AB, internal edge attached to B, C, D, respectively.
The momentum $P$ corresponds to the edge AC, 
and $\Theta _B, \Theta _C, \Theta _D$ are chosen to give a canonical symplectic form, 
such like $\{X_i, P_j\}=\delta _{ij}$.
In this paper, we chose these momenta as 
\begin{equation}
	\Theta _B=y'+w''+y, \Theta _C=x', \Theta _D=y+y'
\end{equation}

We note that this construction is consistent with the tetrahedron with trivial holonomies.
If we tune three of holonomies to be trivial, i.e. $C_B=C_C=C_D=0$,
the holonomy around the last vertex A is automatically set to be trivial 
and we can get back to a tetrahedron with trivial holonomy.

Our polarization is made from the start point via the symplectic transformation like
\begin{equation}
	\begin{pmatrix}
		X \\ C_B \\ C_C \\ C_D \\ P \\ \Theta _B \\ \Theta _C \\ \Theta _D
	\end{pmatrix}
	=\begin{pmatrix}
		x'+z'' \\ w+x'+y'' \\ w''+x''+z' \\ w'+y'+z'' \\ z'+y'' \\ \Theta _B \\ \Theta _C \\ \Theta _D
	\end{pmatrix} 
	=	\begin{pmatrix}
		1 & 0 & 0 & 0 & 0 & 0 & 1 & 0 \\
		0 & 1 & 0 & -1 & 0 & 0 & 1 & -1 \\
		0 & 0 & 1 & 0 & 1 & 1 & 0 & 0 \\
		1 & -1 & 0 & 1 & 0 & -1 & 0 & 0 \\
		0 & 0 & 0 & -1 & 1 & 0 & 0 & -1 \\
		0 & 0 & 0 & 1 & 0 & 1 & 0 & 1 \\
		0 & 0 & 0 & 0 & 0 & 0 & 1 & 0 \\
		0 & 0 & 0 & 1 & 0 & 0 & 0 & 1 \\
	\end{pmatrix}
	\begin{pmatrix}
		z'' \\ w \\ x'' \\ y' \\ z' \\ w'' \\ x' \\ y
	\end{pmatrix}
	+\begin{pmatrix}
		0 \\ i\pi \\ 0 \\ i\pi \\ i\pi \\ 0 \\ 0 \\ 0 
	\end{pmatrix}.
\end{equation}
We will denote this $\Sp (8, \mathbb{Z})$ matrix as M.

According to  \cite{Hua1949}, we can decompose this matrix M into 
generating components of $\Sp(8, \mathbb{Z})$, as
\begin{equation}
	M=USRXJV
\end{equation}
where
\begin{equation}
	\begin{split}
		U=\begin{pmatrix}
			0 & 0 & 1 & 0 & 0 & 0 & 0 & 0\\
			-1 & 0 & 1 & 0 & 0 & 0 & 0 & 0\\
			0 & 1 & 0 & 0 & 0 & 0 & 0 & 0\\
			1 & 0 & 0 & 1 & 0 & 0 & 0 & 0\\
			0 & 0 & 0 & 0 & 1 & 0 & 1 & -1  \\
			0 & 0 & 0 & 0 & -1 & 0 & 0 & 1  \\
			0 & 0 & 0 & 0 & 0 & 1 & 0 & 0  \\
			0 & 0 & 0 & 0 & 0 & 0 & 0 & 1 
		\end{pmatrix}, 
		S=\begin{pmatrix}
			0 & 0 & 0 & 0 & -1 & 0 & 0 & 0  \\
			0 & 0 & 0 & 0 & 0 & -1 & 0 & 0   \\
			0 & 0 & 0 & 0 & 0 & 0 & -1 & 0  \\
			0 & 0 & 0 & 1 & 0 & 0 & 0 & 0 \\
			1 & 0 & 0 & 0 & 0 & 0 & 0 & 0\\
			0 & 1 & 0 & 0 & 0 & 0 & 0 & 0\\
			0 & 0 & 1 & 0 & 0 & 0 & 0 & 0\\
			0 & 0 & 0 & 0 & 0 & 0 & 0 & 1 
		\end{pmatrix}, \\
		R=\begin{pmatrix}
			-1 & 0 & 0 & 1 & 0 & 0 & 0 & 0\\
			0 & -1 & 0 & 0 & 0 & 0 & 0 & 0\\
			0 & 0 & -1 & 0 & 0 & 0 & 0 & 0\\
			-1 & 0 & 0 & 2 & 0 & 0 & 0 & 0\\
			0 & 0 & 0 & 0 & -2 & 0 & 0 & -1  \\
			0 & 0 & 0 & 0 & 0 & -1 & 0 & 1  \\
			0 & 0 & 0 & 0 & 0 & 0 & -1 & 0  \\
			0 & 0 & 0 & 0 & 1 & 0 & 0 & 1 
		\end{pmatrix}, 
		X=\begin{pmatrix}
			1 & 0 & 0 & 0 & 0 & 0 & 0 & 0\\
			0 & 1 & 0 & 0 & 0 & 0 & 0 & 0\\
			0 & 0 & 1 & 0 & 0 & 0 & 0 & 0\\
			0 & 0 & 0 & 1 & 0 & 0 & 0 & 0\\
			0 & 0 & 0 & 0 & 1 & 0 & 0 & 0  \\
			0 & 0 & -1 & 0 & 0 & 1 & 0 & 0   \\
			0 & -1 & 0 & 0 & 0 & 0 & 1 & 0  \\
			0 & 0 & 0 & -1 & 0 & 0 & 0 & 1  
		\end{pmatrix}, \\
		J=\begin{pmatrix}
			0 & 0 & 0 & 0 & -1 & 0 & 0 & 0  \\
			0 & 0 & 0 & 0 & 0 & -1 & 0 & 0   \\
			0 & 0 & 0 & 0 & 0 & 0 & -1 & 0  \\
			0 & 0 & 0 & 0 & 0 & 0 & 0 & -1 \\ 
			1 & 0 & 0 & 0 & 0 & 0 & 0 & 0\\
			0 & 1 & 0 & 0 & 0 & 0 & 0 & 0\\
			0 & 0 & 1 & 0 & 0 & 0 & 0 & 0\\
			0 & 0 & 0 & 1 & 0 & 0 & 0 & 0
		\end{pmatrix}, 
		V=\begin{pmatrix}
			1 & -1 & 0 & 0 & 0 & 0 & 0 & 0\\
			0 & 0 & 1 & 0 & 0 & 0 & 0 & 0\\
			1 & 0 & 0 & 0 & 0 & 0 & 0 & 0\\
			-1 & 1 & 0 & 1 & 0 & 0 & 0 & 0\\
			0 & 0 & 0 & 0 & 0 & -1 & 0 & 1\\
			0 & 0 & 0 & 0 & 0 & 0 & 1 & 0 \\
			0 & 0 & 0 & 0 & 1 & 1 & 0 & 0  \\
			0 & 0 & 0 & 0 & 0 & 0 & 0 & 1 
		\end{pmatrix}.
	\end{split}
\end{equation}
In this decomposition, matrices U, V, R belong to GL-type, S, J to S-type, and X to T-type transformation.
For each type, we can define the corresponding action on the partition functions like $\SL(2,\mathbb{Z})$ case,
which is described in \cite{Dimofte2011}.
We will just refer to their result for $\Sp(2N,\mathbb{Z})$ here;
\begin{eqnarray}
	T:g=\begin{pmatrix}
		I & 0 \\
		B & I
	\end{pmatrix},B =B^{\rm T}
	: Z(\vec{\tilde{m}})& \mapsto& Z'(\vec{\tilde{m}})=e^{i\pi\vec{\tilde{m}}\cdot B\vec{\tilde{m}}} Z(\tilde{m}) \\
	S:g=\begin{pmatrix}
		I-J & -J \\
		J & I-J
	\end{pmatrix}:
	Z(\tilde{m})& \mapsto& Z'(\vec{\tilde{m}'})=\int d\tilde{m}
	e^{-2i\pi \vec{\tilde{m}}\cdot J\vec{\tilde{m}'}} Z(\tilde{m}),
\end{eqnarray}
where $J = {\rm diag}(j_1 ,\cdots, j_N)$ with $j_i \in \{0,1\}$ and
the integration is over i-th component of $\vec{m}$ with $j_i=1$, and
\begin{equation}
	GL:g=\begin{pmatrix}
		U & 0 \\
		0 & U^{{\rm T}  -1}
	\end{pmatrix}, U \in \GL(N,\mathbb{Z}):
	Z(\vec{\tilde{m}}) \mapsto Z'(\vec{\tilde{m}}')=Z(U^{-1}\vec{\tilde{m}}').
\end{equation}
There can be transformations of the constant shift in units of $i\pi$.
We will call these transformations affine shifts.
When an affine shift occurs to the coordinate as 
$\begin{pmatrix}
	X' \\ P'
\end{pmatrix}=\begin{pmatrix}
	X \\ P
\end{pmatrix}+\begin{pmatrix}
	i\pi \\ 0
\end{pmatrix}$,
the partition function changes as
\begin{equation}
	Z(X)\mapsto Z'(X)=Z(X-\frac{iQ}{2}).
\end{equation}
When it occurs to the momentum, we can get 
$\begin{pmatrix}
	X' \\ P'
\end{pmatrix}=\begin{pmatrix}
	X \\ P
\end{pmatrix}+\begin{pmatrix}
	0 \\ i\pi
\end{pmatrix}=
-S\left(S\begin{pmatrix}
	X \\ P
\end{pmatrix}+\begin{pmatrix}
	i\pi \\ 0
\end{pmatrix}\right)$,
where $S=\begin{pmatrix}
	0 & -1 \\
	1 & 0
\end{pmatrix}$ is a S-type transformation.
If we combine these transformations we can get
\begin{equation}
	\begin{split}
		Z(X)\mapsto Z'(X'')&=\int dXdX' e^{-2i\pi(-X'')X'-2i\pi X(X'-\frac{iQ}{2})}Z(X) \\
		&=e^{\pi QX''}Z(X'').
	\end{split}
\end{equation}

Now we are ready to compute the partition function of the tetrahedron with three nontrivial holonomies.
First we prepare the partition function 
\begin{equation}
	Z(z'', w, x'', y')=Z_{\Delta}(z'')Z_{\Delta}(w)Z_{\Delta}(x'')Z_{\Delta}(y'), 
\end{equation}
and then act a symplectic transformation $M$ on $Z(z'', w, x'', y')$.
The practical computation is tedious, so we only give a result here;
\begin{equation}
	\begin{split}
		Z^M (X;C_B, C_C, C_D)=\int dz_1 e^{-2i\pi z_1 \left( X-C_B+\frac{iQ}{2}\right) -2i\pi z_1 ^2
	-i\pi\left(iQ-X\right)^2-\pi Q\left(iQ-X\right)} \\
	\times e^{-2i\pi C_C\left( X+z_1 - \frac{iQ}{2}\right)- i\pi C_C^2
+\pi Q\left(C_B+C_D\right)} \\
\times Z_{\Delta}(X)Z_{\Delta}(z_1+C_C)Z_{\Delta}(iQ-X-z_1-C_C) \\
\times Z_{\Delta}(z_1+X-C_B-C_D)Z_{\Delta}(-z_1).
\end{split}
\end{equation}
We propose this final result as a partition function of the solid pants.
The $z_1$ integration is too hard to compute, so it remains at the last result.
The change of the number of $Z_{\Delta}$ is caused by pentagon identity.

We can check this calculation in two ways.
First, there is a rotational symmetry which exchanges $C_B, C_C, C_D$ combining with 
$ST$ transformation over $X$.
We will name this rotation $Q$.
Geometrically $Q$ is a remnant of $ST$ symmetry of a single tetrahedron.
Actually $Q$ can be constructed from $M$ and a new transformation $R_1$.
The transformation $R_1$ acts on the four tetrahedra before glued, as
changing the order of $x, y, z$, and rotating $w, w', w''$ at the same time.
If we represent it as a symplectic transformation, then we get
\begin{equation}
	\begin{pmatrix}
		z'' \\ w \\ x'' \\ y' \\ z' \\ w'' \\ x' \\ y
	\end{pmatrix}
	\mapsto
	\begin{pmatrix}
		x'' \\ w' \\ y'' \\ z' \\ x' \\ w \\ y' \\ z
	\end{pmatrix}
	=\begin{pmatrix}
		0 & 0 & 1 & 0 & 0 & 0 & 0 & 0 \\
		0 & -1 & 0 & 0 & 0 & -1 & 0 & 0 \\
		0 & 0 & 0 & -1 & 0 & 0 & 0 & -1 \\
		0 & 0 & 0 & 0 & 1 & 0 & 0 & 0 \\
		0 & 0 & 0 & 0 & 0 & 0 & 1 & 0 \\
		0 & 1 & 0 & 0 & 0 & 0 & 0 & 0 \\
		0 & 0 & 0 & 1 & 0 & 0 & 0 & 0 \\
		-1 & 0 & 0 & 0 & -1 & 0 & 0 & 0 
	\end{pmatrix}
	\begin{pmatrix}
		z'' \\ w \\ x'' \\ y' \\ z' \\ w'' \\ x' \\ y
	\end{pmatrix}
	+
	\begin{pmatrix}
		0 \\ -i\pi \\ i\pi \\ 0 \\ 0 \\ 0 \\ 0 \\ i\pi	
	\end{pmatrix}.
\end{equation}
Then if we substitute $Q = M R_1 M^{-1}$, 
the action of $Q$ over the polarization  
$(X, C_B, C_C, C_D$ , $P, \Theta _B, \Theta _C, \Theta _D)^{\rm T}$ looks as a combination of
an $ST$ transformation over $(X, P)$ and a rotation of $(C_B, C_C, C_D)$.

The second check is setting $C_B=C_C=C_D=0$.
This means setting all the holonomies to be trivial, and in fact
remaining integration becomes executable as
\begin{equation}
	\begin{split}
		Z^M (X;0, 0, 0)&=\int dz_1 e^{-i\pi z_1 ^2 - i\pi \left( z_1 +X\right )^2 -\pi Q \left(z_1+X\right)}
		Z_{\Delta}(X)Z_{\Delta}(z_1)Z_{\Delta}(iQ-X-z_1)Z_{\Delta}(-z_1)Z_{\Delta}(z_1+X) \\
		&= \int dz_1 e^{-i\pi z_1 ^2 +\pi Qz_1}Z_{\Delta}(X)Z_{\Delta}(-z_1)Z_{\Delta}(z_1) \\
		&\propto Z_{\Delta}(X).
	\end{split}
	\label{}
\end{equation}
In a result $Z^M(X)$ becomes equal to $Z_{\Delta}(X)$ up to constant coefficient, 
and we find that a tetrahedron with nontrivial holonomies reduces to that with trivial holonomies.

\section{Discussion}

We proposed a method to compute holonomy matrices for hyperbolic 3-manifolds with boundaries.
By using this method, we analyzed the hyperbolic structure of an ideal tetrahedron 
and generalized to have nontrivial holonomies.
And we calculated a partition function of the nontrivial tetrahedron via the duality \cite{Dimofte2011}, 
and check some consistencies.
Our proposing partition function still containing the form of integration, 
so that the dual 3d gauge theory is mysterious.

\paragraph{Acknowledgement}
I would like thank Y. Matsuo and  my colleagues for valuable discussions and comments.

\appendix
\section{Formula for double sine functions}

In this section we show some formula of partition functions used in Section 4.
We derive these formula from the formulas of quantum dilogarithm functions referring to 
\cite{Barnes1899, Faddeev2000}.

The quantum dilogarithm function is defined be the formula
\begin{equation}
	e_b(z)=\exp\left(\frac{1}{4}\int _{-\infty}^{\infty} \frac{e^{-2izx}dx}{\sinh(xb)\sinh(xb^{-1})x}\right), 
\end{equation}
where the integration go beyond the singularity at $x=0$. 
The double sine function is the function defined as
\begin{equation}
	s_b(z)=\prod_{m, n \in \mathbb{Z}_{\geq 0}}\frac{mb+nb^{-1}+\frac{Q}{2}-ix}{mb+nb^{-1}+\frac{Q}{2}+ix}, 
\end{equation}
where $Q=b+b^{-1}$.
Actually these special functions are related as 
\begin{equation}
	e_b(z)=e^{\frac{iQ}{2}z^2}s_b(z), 
\end{equation}
thus the partition function which is dual to an ideal tetrahedron can be written as 
\begin{equation}
	\begin{split}
		Z_{\Delta}(z)&=e^{\frac{i\pi}{2}\left(\frac{iQ}{2}-z\right)^2}s_b\left(\frac{iQ}{2}-z\right) \\
		&=e_b\left(\frac{iQ}{2}-z\right).
	\end{split}
\end{equation}

Quantum dilogarithm function has following properties.
\begin{eqnarray}
	e_b(z)e_b(-z)=e^{i\pi z^2-\frac{i\pi}{6} \left( 1-\frac{Q^2}{2}\right) } \hspace{5cm}\\
	\int dx e_b(x)e^{2i\pi y}=
	e^{-i\pi y^2+\frac{i\pi}{12}\left(1+Q^2\right)}e_b\left(y+\frac{iQ}{2}\right) \hspace{3cm}\\
	e_b\left(x+\frac{iQ}{2}\right)e_b\left(y+\frac{iQ}{2}\right)e^{2i\pi xy} \nonumber \hspace{9cm}\\
	= \int dz e_b\left(z+\frac{iQ}{2}\right)e_b\left(x-z+\frac{iQ}{2}\right)e_b\left(y-z+\frac{iQ}{2}\right) 
	e^{-2i\pi z^2 + 2i\pi z(x+y) +\frac{i\pi}{12}\left(1+Q^2\right)} \\
	e_b\left(x+\frac{iQ}{2}\right)e_b\left(\frac{iQ}{2}+u-x\right)e_b(-u-\frac{iQ}{2})
	e^{-i\pi u^2 +\pi Q u}  \nonumber \hspace{5cm}\\
	= \int dz e_b\left(z+\frac{iQ}{2}\right)e_b\left(x-z+\frac{iQ}{2}\right)
	e^{-i\pi z^2 - 2i\pi z\left(\frac{iQ}{2}+u-x\right) -\frac{i\pi}{12}\left(1+Q^2\right)}, 
\end{eqnarray}
where the integration is over $z \in \mathbb{R}$ and singularities are put below except for at $z=0$.
The third and fourth relations are called pentagon identities.
When we cast these identities into the word of the partition functions, we get
\begin{eqnarray}
	Z_{\Delta}(z+iQ)Z_{\Delta}(-z)=e^{i\pi z^2 -\pi Qz -\frac{iQ}{6}\left(1+Q^2\right)} \hspace{4cm}\\
	\int dz Z_{\Delta}(z) e^{2i\pi zw}=-e^{-i\pi w^2 -\pi Qw + \frac{iQ}{12}\left(1+Q^2\right)}Z_{\Delta}(w)
	\hspace{2cm}\\
	Z_{\Delta}(x)Z_{\Delta}(y) e^{2i\pi xy} \hspace{10cm}\nonumber \\
	= \int dz Z_{\Delta}(-z)Z_{\Delta}(z-x)Z_{\Delta}(z-y) 
	e^{-2i\pi z^2 + 2i\pi z(x+y)+\frac{iQ}{12}\left(1+Q^2\right)} \\
	Z_{\Delta}(iQ-u)Z_{\Delta}(x)Z_{\Delta}(u-x) e^{-i\pi u^2 -\pi Qu} \hspace{7cm}\nonumber\\
	=	\int dz Z_{\Delta}(-z)Z_{\Delta}(z+x)
	e^{-i\pi z^2 -2i\pi \left(\frac{iQ}{2}-u+x\right)-\frac{i\pi}{12}\left(1+Q^2\right)}\hspace{2cm}
\end{eqnarray}

\bibliography{Collection,Add_Collection}

\end{document}